\documentstyle[prl,aps]{revtex}
\tighten
\begin{document}

\draft
\title{Nucleosynthesis in Supernovae}

\author{C. J. Horowitz\footnote{email: charlie@iucf.indiana.edu} 
and Gang Li\footnote{email: ganli@indiana.edu}} 
\address{Nuclear Theory Center\\
2401 Milo B. Sampson Lane \\
Bloomington, Indiana 47405}

\date{\today} 
\maketitle 
\begin{abstract}
Core collapse supernovae are dominated by energy transport from
neutrinos.  Therefore, some supernova properties could depend
on symetries and features of the standard model weak interactions.  The cross section
for neutrino capture is larger than that for antineutrino capture by one term 
of order the neutrino energy over the nucleon mass.  This reduces the ratio of 
neutrons to protons in the $\nu$-driven wind above a protoneutron star 
by approximately 20 \% and may significantly hinder r-process nucleosynthesis.
\end{abstract}
\pacs{95.30.Cq, 26.30.+k,11.30.Er, 97.60.Bw}

Core collapse supernovae are perhaps the only present day large systems
dominated by the weak interaction.   They are so dense that photons and
charged particles diffuse very slowly.   Therefore energy transport is by
neutrinos (and convection).

We beleive it may be useful to try and relate some supernova properties 
to the symmetries and features of the standard model weak interaction. 
Parity violation in a strong magnetic field could lead to an asymmetry of
the explosion\cite{1}.  Indeed, supernovae explode with a dipole asymmetry of
order one percent in order to produce the very high `recoil' velocities
observed for  neutron stars\cite{2}.   However, calculating the expected
asymmetry from P violation has proved complicated.  Although explicit
calculations have yielded somewhat small asymmetries\cite{3,4,5} it is still
possible that more efficient mechanisms will be found.

In this letter we calculate some effects from the difference between 
neutrino and antineutrino interactions.  In  Quantum Electrodynamics 
the cross section for $e^-p$ is equal to that for $e^+p$ scattering 
(to lowest order in $\alpha$).  In contrast, the standard model has
$\bar\nu$-nucleon cross sections systematically smaller than 
$\nu$-nucleon cross sections.

However at the low $\nu$ energies in supernovae, time reversal symmetry
limits the difference between $\nu$ and $\bar\nu$ cross sections.  
Time reversal can relate $\nu-N$ elastic scattering and $\bar\nu-N$ where 
the nucleon scatters from final momentum $p_f$ to initial momentum $p_i$.  
If the nucleon does not recoil then the $\nu$ and $\bar\nu$ cross sections 
are equal.   Thus the difference between $\nu$ and $\bar\nu$ cross sections 
are expected to be of recoil order $E/M$ where $E$ is the neutrino energy 
and $M$ the nucleon mass.
\footnote{We expect the difference for charged
current interactions to be of the same order if one can neglect the
neutron-proton mass difference.}   This ratio is relatively small in
supernovae.  However the coefficient multiplying $E/M$ involves the large
weak magnetic moment of the nucleon (see below).

The standard model has larger $\nu$ cross sections than those for $\bar\nu$.
For neutral currents, this leads to a longer mean free path for $\bar\nu_x$ 
compared to $\nu_x$ (with x=$\mu$ or $\tau$).  Thus even 
though $\nu_x$ and $\bar\nu_x$ are produced in pairs, the antineutrinos 
escape faster leaving the star neutrino rich.  The muon and tau number 
for the protoneutron star in a supernova could be of order $10^{54}$\cite{6}.  
Supernovae may be the only known systems with large $\mu$ and or $\tau$ number.
For charged currents, the interaction difference can change the 
equilibrium ratio of neutrons to protons and may have important 
implications for nucleosynthesis.  We discuss this below.  To our knowledge, 
all previous work on nucleosynthesis in supernovae assumed equal $\nu$ 
and $\bar\nu$ interactions (aside from the n-p mass difference).

The neutrino driven wind outside of a protoneutron star is an attractive
site for r-process nucleosynthesis\cite{7}.  Here nuclei rapidly capture neutrons
from a low density medium to produce heavy elements\cite{8}.    This requires,
as a bare minimum, that the initial material have more neutrons than protons.
The ratio of neutrons to protons n/p in the wind depends on the rates for
the two reactions:
$$\nu_e + n \rightarrow p + e^-,\eqno(1a)$$
$$\bar\nu_e+p \rightarrow n + e^+.\eqno(1b)$$
The standard model cross sections for Eqs. (1a,1b) to order $E/M$ are,
$$\sigma={G^2{\rm cos}^2\theta_c\over \pi}(1+3g_a^2) E_e^2
[1-\gamma {E\over M} \pm \delta{E\over M}],\eqno(2)$$
with $G$ the Fermi constant (and $\theta_c$ the Cabbibo angle),
$E_e=E\pm \Delta$ the energy of the charged lepton and $\Delta=1.293$ MeV is
the neutron-proton mass difference.  The plus sign is for Eq. (1a) and the
minus sign for Eq. (1b).  We use $g_a\approx 1.26$.\footnote{ Note in
principle, there is another correction to Eq. (2) from the thermal motion of
the nucleons.  This is of order $T/M$ and increases both the $\nu$ and
$\bar\nu$ cross sections.  However we assume the temperature in the wind $T$
is much less than the neutrino sphere temperature\cite{11} and neglect this term.}
Equation (2) neglects small corrections involving the electron mass and coulomb 
effects (see for example\cite{engel}) while the finite nucleon size only enters 
at order $(E/M)^2$.

We refer to the $\gamma$ term as a recoil correction.  It is the same for
$\nu$ and $\bar\nu$.
$$\gamma=(2+10g_a^2)/(1+3g_a^2)\approx 3.10\eqno(3)$$
Finally, the $\delta$ term involves the interference of vector (1+2$F_2$) and 
axial ($g_a$) currents.  This violates P, which by CP invariance also violates
C.   This increases the $\nu$ and decreases the $\bar\nu$ cross section.
$$\delta=4g_a(1+2F_2)/(1+3g_a^2)\approx 4.12\eqno(4)$$
Here $F_2$ is the isovector anomalous moment of the nucleon. (This is the
weak magnetism contribution.)

We average Eq. (2) over the $\nu_e$ spectrum to get,
$${\rm <}\sigma{\rm >}_\nu={G^2{\rm cos}^2\theta_c\over
\pi}(1+3g_a^2) {\rm <}E{\rm >}\epsilon [1+2{\Delta\over \epsilon}+ a_0
{\Delta^2\over \epsilon^2}] [1+(\delta-\gamma)a_2 {\epsilon\over M}],
\eqno(5)$$
for Eq. (1a).  Here the mean energy $\epsilon$ is defined as,
$$\epsilon={\rm <}E^2{\rm >}/{\rm <}E{\rm >},\eqno(6)$$
and $a_2$ is a shape factor $a_2={\rm <}E^3{\rm >< }E{\rm >}/{\rm
<} E^2 {\rm >}^2$.  Finally $a_0={\rm <}E^2{\rm>}/{\rm <} E{\rm >}^2$ and
${\rm <}E^i{\rm >}$ are the ith energy moments of the $\nu_e$ spectrum.
Note, $\epsilon\approx 1.2 {\rm <}E{\rm >}$.

Likewise, averaging over the $\bar\nu_e$ spectrum for Eq. (1b) gives,
$${\rm <}\sigma{\rm >}_{\bar\nu}={G^2{\rm cos}^2\theta_c\over \pi}(1+3g_a^2)
{\rm <}\bar E{\rm >}\bar\epsilon [1-2{\Delta\over \bar\epsilon}+ a_0
{\Delta^2\over \bar\epsilon^2}] [1-(\delta+\gamma)a_2 {\bar\epsilon\over M}],
\eqno(7)$$
with the mean antineutrino energy $\bar\epsilon={\rm<}\bar E^2{\rm >}/{\rm
<} \bar E {\rm >}$ and  ${\rm <}\bar E^i{\rm >}$ the ith moment of the
$\bar\nu_e$ spectrum.   We assume similar shape factors $a_2$ and $a_0$ for
$\bar\nu_e$ and $\nu_e$.  The shape factor $a_2=1.23$ (1.15) for a Fermi
Dirac distribution with chemical potential $\mu=\eta T_\nu$ and temperature
$T_\nu$ for $\eta=0$ (3.5).  See for example\cite{9}.  For simplicity we adopt
$a_2=a_0=1.2$
\footnote{The coefficient $a_0$ only makes a very small contribution
and our results are insensitive to its value.}.

The equilibrium electron fraction per baryon $Y_e$ (which is equal to the
proton fraction assuming charge neutrality) is simply related to the rate
$\bar\lambda$ for Eq. (1b) divided by the rate $\lambda$ for Eq. (1a).
$$Y_e=(1+{\bar\lambda\over \lambda})^{-1}\eqno(8a)$$
This assumes the neutrino capture rates dominate those 
for other reactions.  Reference\cite{10} contains some discussion of the small
corrections from $e^\pm$ capture.  The ratio n/p is,
$${n\over p} = {1\over Y_e} -1.\eqno(8b)$$
Taking the ratio of Eq. (7) to Eq. (5) gives,
$$Y_e=\Bigl(1+{L_{\bar\nu_e}\bar\epsilon\over L_{\nu_e}\epsilon}
QC\Bigr)^{-1}.\eqno(9)$$
Here $L_{\nu_e}$ ($L_{\bar\nu_e}$) is the $\nu_e$ ($\bar\nu_e$) luminosity,
$Q$ is the correction from the reaction Q value,
$$Q={1-2{\Delta\over \bar\epsilon}+a_0{\Delta^2\over \bar\epsilon^2}\over
1+2{\Delta\over \epsilon}+a_0{\Delta^2\over\epsilon^2}},\eqno(10)$$
and the C violating term, Eq. (4), contributes the factor $C$,
$$C={1-(\delta+\gamma)a_2{\bar\epsilon\over M}\over
1+(\delta-\gamma)a_2{\epsilon\over M}}.\eqno(11)$$
Note, the recoil term $\gamma$ makes a small but nonzero contribution to
Eq. (11) because the $\nu$ and $\bar\nu$ energies are different.  
Simply evaluating Eq. (11) for typical parameters yields $C\approx 0.8$.  
Thus, {\it the difference between $\nu$ and $\bar\nu$ interactions reduces the 
equilibrium n/p ratio by approximately 20 \%.}  This is a major result of the present 
paper and will be discussed below.

Figure 1 shows the values of $\epsilon$ and $\bar\epsilon$ necessary for
$Y_e=0.5$.  We assume equal luminosities $L_{\nu_e}=L_{\bar\nu_e}$.  The
region to the upper left is neutron rich and to the lower right proton rich.
The conditions for $Y_e=0.5$, assuming $C=1$ in Eq. (9), are indicated by the 
dotted line.  Including $C$ shifts the conditions for $Y_e=0.5$ to the solid 
line.  Thus the difference in $\nu$ and $\bar\nu$ interactions converts the 
region between the solid and dotted lines from neutron rich to proton rich.

We also show in Fig. 1 the values of $\epsilon$ and $\bar\epsilon$ from a
supernova simulation by J.R. Wilson as reported in ref.\cite{11}.  The
symbols show how the mean energies evolve with time.  As the protoneutron
star becomes more neutron rich, the opacity for $\bar\nu_e$ decreases because
there are fewer protons.  This allows the $\bar\nu_e$ to escape from deeper
inside the hot protoneutron star.  Therefore $\bar\epsilon$ increases with
time.   Without $C$ the wind starts out with  $Y_e\approx 0.5$ and then 
becomes neutron rich.  With $C$ the wind starts out proton rich and 
ends up with $Y_e\approx 0.5$.  If $L_{\bar\nu_e}\approx L_{\nu_e}$ 
the wind is never significantly neutron rich.  
If $L_{\bar\nu_e}\approx 1.1 L_{\nu_e}$ the wind will end 
slightly neutron rich.  However, n/p is still 20 \% lower with $C$  
than without.  For example, if $Y_e$ drops as low as 0.42 in a model 
without $C$ it will only drop to approximately 0.48 when the difference 
between $\nu$ and $\bar\nu$ interactions is included.  With this increase in
$Y_e$, it is very unlikely that successful r-process nucleosynthesis can
take place in the wind of this or similar models.

Note, we are being slightly inconsistent to include the $E/M$ term in Eqs. (2,4) for 
the neutrino absorption while it is not included in the simulation used 
for $\epsilon$ and $\bar\epsilon$.  Indeed this term could change
the location of the neutrino spheres and {\it slightly} increase
$\bar\epsilon$ and decrease $\epsilon$.  This could cancel a small part of
the effect on the n/p ratio.  However, our preliminary estimates suggest
this change in the spectrum is very small.  Including the term 
in a full simulation would be useful\cite{12}.  For completeness we give a 
C violating term for neutrino-electron scattering NES which may be useful for
calculating differences between the $\nu_x$ and $\bar\nu_x$ spectrum.

The total cross section $\sigma_e$ for NES (see ref.\cite{13} for example) is
expanded in powers of $E/E_F$ where $E_F$ is the electron Fermi energy.  To
order $(E/E_F)^2$,
$$\sigma_e\approx {G^2E^2\over \pi}(c_v^2+c_a^2){E\over 5E_F}\bigl(1\pm
\delta_e{E\over E_F}\bigr),\eqno(12)$$
with $\delta_e=4c_vc_a/3(c_v^2+c_a^2)$ and the plus sign is for $\nu$ and
the minus sign for $\bar\nu$.  The couplings are 
$c_v=2{\rm sin}^2\theta_W\pm1/2$ and $c_a=\pm1/2$.  Here the plus sign is 
for $\nu_e$ and the minus sign for $\nu_x$. The C violating coefficient
$\delta_e\approx 0.55$ for $\nu_e$ and $\approx 0.1$ for $\nu_x$.
Although this term is nominally of larger order, $E/E_F$ for 
NES than $E/M$ for nucleon scattering, the coefficient is smaller $\delta_e\ll
\delta$.  Therefore we do not expect large differences from NES
(except perhaps at low densities).

With the approximately 20 \% reduction in n/p from the difference 
between $\nu$ and $\bar\nu$ interactions, there appears to be 
very serious problems with r-process nucleosynthesis in the
wind of present supernova models.  In addition to the initial lack of
neutrons,  one has to overcome the effects of neutrino interactions during
the assembly of $\alpha$ particles and during the r-process itself\cite{14}.
These further limit the available neutrons per seed nucleus.  Thus, it is
unlikely that present wind models will produce a successful r-process.  Of
course, the wind in supernovae may not be the r-process site, although this
may be unappealing (see for example\cite{8,15}).  If the wind is not the site,
one must look for alternative environments.

However, the effects of neutrino interactions may be very general.  The only 
requirement is that energy transport from neutrinos plays some role in helping 
material out of a deep gravitational well.   Given this, it is quite likely 
that the n/p ratio will be determined by the relative rates of Eqs. (1a,1b).  
Therefore differences in $\nu$ and $\bar\nu$ interactions may be important for 
just about any nucleosynthesis site that involves neutrinos.  Indeed, 
Haxton et al.\cite{16} claim the abundance of isotopes produced by 
neutrino spallation imply significant neutrino fluences during the r-process.

If the $\nu$-driven wind is the r-process site, it is very likely,
present models of the neutrino radiation in supernovae are incomplete.  The
high values of $Y_e$ make it almost impossible to have a successful r-process
by only changing matter properties, such as the entropy.  The neutrino fluxes
will (almost assuredly) need to be changed.

Changes in the astrophysics used in the simulations or new neutrino physics
such as neutrino oscillations\cite{17} could change $\bar\epsilon$, $\epsilon$
and or the luminosities and lead to a more neutron rich wind.  The
oscillations of more energetic $\bar\nu_x$ with $\bar\nu_e$ could increase
$\bar\epsilon$.  However, we have some information on the $\bar\nu_e$
spectrum from SN1987a\cite{18}.  Thus one can not increase $\bar\epsilon$ 
without limit.  Indeed if anything, the Kamiokande data suggest a lower
$\bar\epsilon$.  Any model which tries to solve r-process nucleosynthesis
problems by increasing $\bar\epsilon$ should first check consistency with
SN1987a observations\cite{19}.  Alternative modifications could include
oscillations of $\nu_e$ to a sterile neutrino or a {\it lowering} of
$\epsilon$.  (However, we know of no model which lowers $\epsilon$.)
Whatever the modification of the neutrino fluxes, one will still need to
include the differences between $\nu$ and $\bar\nu$ interactions
in order to accurately calculate n/p.

In conclusion, supernovae are one of the few large systems dominated by
energy transport from weakly interacting neutrinos.  Therefore, some 
supernova properties may depend on symmetries and features
of the standard model weak interactions.  The cross secton for neutrino
capture is larger than that for antineutrino capture by a term of
order the neutrino energy over the nucleon mass.  This difference between 
neutrino and antineutrino interactions reduces the ratio of neutrons 
to protons in the $\nu$-driven wind above a protoneutron star by 
approximately 20 \% and may significantly hinder r-process 
nucleosynthesis.

This work was supported in part by DOE grant: DE-FG02-87ER40365.

\eject


\vbox to 4.in{\vss\hbox to 8in{\hss
{\includegraphics{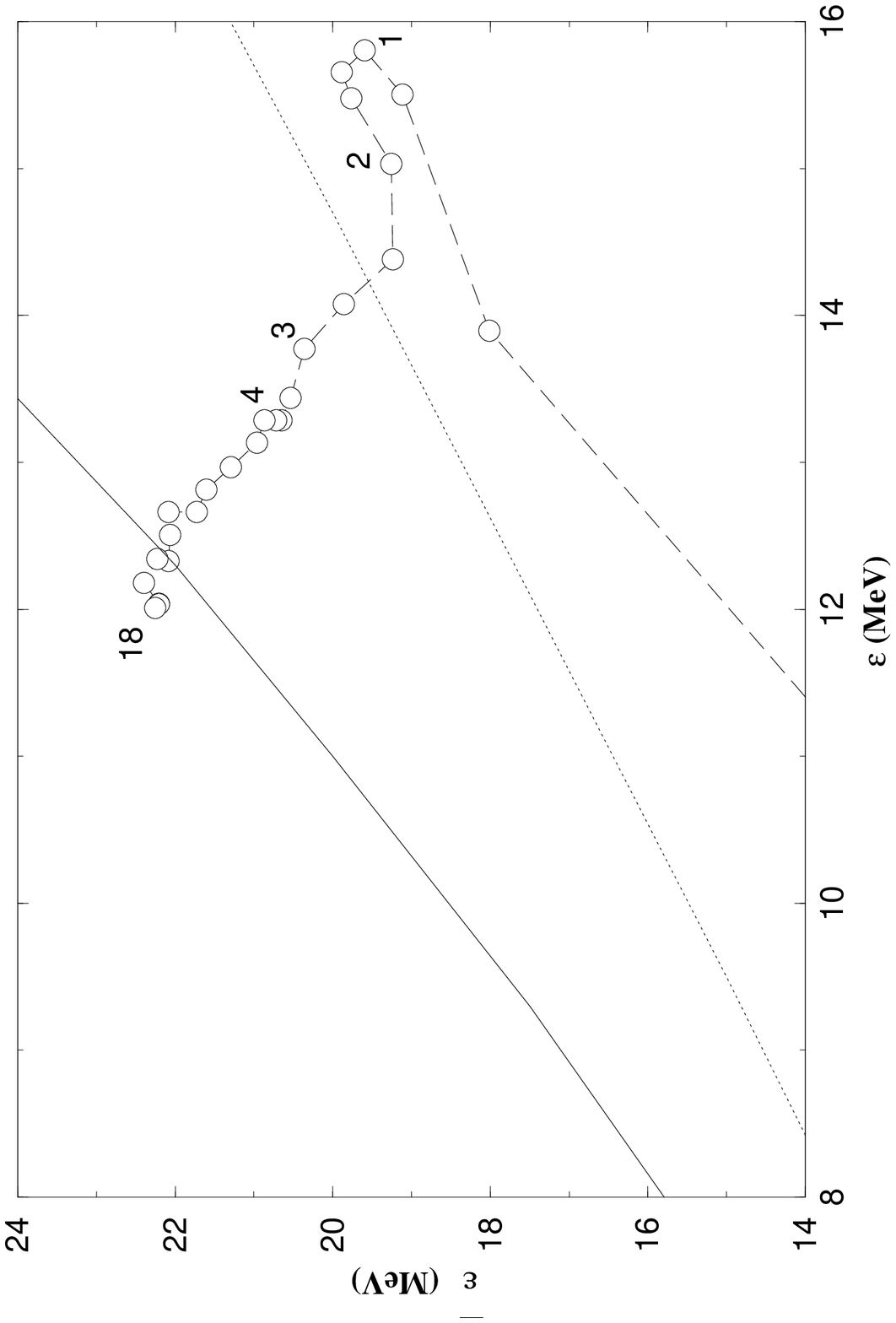}}\hss}}
\nobreak
{\noindent\narrower{{\bf FIG.~1}. 
Mean antineutrino energy $\bar\epsilon$ vs mean neutrino energy $\epsilon$, 
see Eq. (6).  The solid line indicates an equilibrium electron fraction 
$Y_e=0.5$ including the difference between $\nu$ and $\bar\nu$ interactions, 
$C$ term in Eqs. (9,11), while the dotted line shows $Y_e=0.5$ without this term.  
The symbols are the mean energies of a simulation by J.R. Wilson as reported 
in ref.\cite{11}  for the indicated times in seconds after collapse.   
The $\nu$-driven wind is neutron rich in the upper left of the figure and 
proton rich in the lower right. The region between the dotted and solid lines 
is converted from neutron rich to proton rich by the $C$ term.}}

\end{document}